\documentclass[aps,pra,twocolumn,superscriptaddress]{revtex4-2}
\usepackage{bbm}
\usepackage{mathrsfs}
\usepackage{amsmath}
\usepackage{amsfonts}
\usepackage[colorlinks=true,linkcolor=blue,urlcolor=blue,citecolor=blue,anchorcolor=blue]{hyperref}
\usepackage{graphicx,epstopdf}
\usepackage{subfigure}
\usepackage{epsfig}
\usepackage{dcolumn}
\usepackage{bm}
\usepackage{color}
\usepackage{natbib}
\usepackage{amssymb}
\usepackage{xcolor}
\usepackage{braket}
\usepackage{ulem}
\usepackage{float}
\usepackage{lipsum}
\usepackage{pifont}

\definecolor{newtxtcolor1}{rgb}{0.8, 0, 0.2}
\definecolor{newtxtcolor2}{rgb}{0.8, 0, 0}

\begin{document}

\title{Manipulating Spectral Windings and Skin Modes through Nonconservative Couplings}

\author{Ningxin Kong}
\affiliation{State Key Laboratory for Mesoscopic Physics, School of Physics, Frontiers Science Center for Nano-optoelectronics, $\&$ Collaborative Innovation Center of Quantum Matter, Peking University, Beijing 100871, China}

\author{Mingsheng Tian}
\thanks{Department of Physics, The Pennsylvania State University, University Park, Pennsylvania, 16802, USA}
\affiliation{State Key Laboratory for Mesoscopic Physics, School of Physics, Frontiers Science Center for Nano-optoelectronics, $\&$ Collaborative Innovation Center of Quantum Matter, Peking University, Beijing 100871, China}

\author{Chenghe Yu}
\affiliation{State Key Laboratory for Mesoscopic Physics, School of Physics, Frontiers Science Center for Nano-optoelectronics, $\&$ Collaborative Innovation Center of Quantum Matter, Peking University, Beijing 100871, China}
\affiliation{Hefei National Laboratory, Hefei 230088, China}

\author{Yilun Xu}
\affiliation{State Key Laboratory for Mesoscopic Physics, School of Physics, Frontiers Science Center for Nano-optoelectronics, $\&$ Collaborative Innovation Center of Quantum Matter, Peking University, Beijing 100871, China}
\affiliation{Beijing Academy of Quantum Information Sciences, Beijing 100193, China}

\author{Matteo Fadel}	
\affiliation{Department of Physics, ETH Z\"{urich}, 8093 Z\"{urich}, Switzerland}

\author{Xinyao Huang}
\email{xinyaohuang@buaa.edu.cn}
\affiliation{School of Physics, Beihang University, Beijing 100191, China}

\author{Qiongyi He}
\affiliation{State Key Laboratory for Mesoscopic Physics, School of Physics, Frontiers Science Center for Nano-optoelectronics, $\&$ Collaborative Innovation Center of Quantum Matter, Peking University, Beijing 100871, China}
 \affiliation{Hefei National Laboratory, Hefei 230088, China}
\affiliation{Collaborative Innovation Center of Extreme Optics, Shanxi University, Taiyuan, Shanxi 030006, China}
\affiliation{Peking University Yangtze Delta Institute of Optoelectronics, Nantong 226010, Jiangsu, China} 

\begin{abstract}
The discovery of the non-Hermitian skin effect (NHSE) has revolutionized our understanding of wave propagation in non-Hermitian systems, highlighting unexpected localization effects beyond conventional theories.
Here, we discover that NHSE, accompanied by multi-type spectral phases, can be induced by manipulating nonconservative couplings.
By characterizing the spectra through the windings of the energy bands, we demonstrate that band structures with identical, opposite, and even twisted windings can be achieved. 
These inequivalent types of spectra originate from the multi-channel interference resulting from the interplay between conservative and nonconservative couplings.
Associated with the multi-type spectra, unipolar and bipolar NHSE with different eigenmode localizations can be observed.
 Additionally, our findings link the nonreciprocal transmission properties of the system to multiple spectral phases, indicating a connection with the skin modes. 
 This paper paves new pathways for investigating non-Hermitian topological effects and manipulating nonreciprocal energy flow.
\end{abstract}
    
    \maketitle
    
\section{Introduction}
Recent advancements in the study of non-Hermitian systems have garnered extensive interest due to their unique attributes and potential applications that extend beyond the domain of traditional Hermitian physics~\cite{Bender2007,Gong2018,Ashida2020,Bergholtz2021,Okuma2023}. Non-Hermitian Hamiltonians, typically employed to describe systems that exchange energy with their environment, have revealed a number of intriguing phenomena~\cite{Liu2016,Xu2016,El-Ganainy2018,Ozdemir2019,Li2023,Li2023a,Wu2022}. 
Among these, the non-Hermitian skin effect (NHSE) results in localization of the eigenstates at the edges of the system, challenging conventional bulk-boundary correspondence and having profound theoretical and experimental implications~\cite{Yao2018,Liang2022,Zhang2022a,YangGBZ2020,Wang2022,Li2020,Lin2023}.
In fact, characteristic signatures of the NHSE are directly reflected in the system's spectrum, which displays complex band structures forming closed loops with nonzero windings in the quasi-energy space under periodic boundary conditions (PBC), distinctly different from those observed under open boundary conditions (OBC)~\cite{Zhang2020_wn,Ding2022}.
Experimentally, the NHSE has been demonstrated in various practical setups including optical systems~\cite{Li2022,Xiao2020}, acoustic  systems~\cite{Zhang2021,Zhang2021a}, and topolectrical circuits~\cite{Helbig2020,Zou2021}. These studies have potential applications in directional amplifiers~\cite{Wen2022,tian23}, enhanced sensors~\cite{McDonald2020,Budich2020}, and efficient energy harvesting~\cite{Xue2022}, showcasing the real-world utility of this effect. 

Recent studies extend the exploration of NHSE by delving into the phenomenon of bipolar NHSE, where eigenmode localization is observed on both sides of a system~\cite{Song2019,Xin2023}. Unlike traditional unipolar NHSE, characterized by the same band windings with one-sided eigenmode localization, bipolar NHSE exhibits multi-type spectral phases associated with diverse eigenmode localization patterns~\cite{Tian2023}. It can manifest as separate loops with opposite windings in multi-band models~\cite{Lin2024,Liu2023} or as twisted loops in single-band scenarios~\cite{Yuce2024,Li2024}, attracting growing research interest in exploring the rich spectral topology of non-Hermitian systems and enlightening the applications in nonreciprocal wave manipulations.

Traditionally, the realization of bipolar NHSE necessitates asymmetric coupling strengths that extend beyond nearest-neighbor couplings~\cite{Wang2021,Wang2021nat,Islam2024,Soumya2024}.
Here, we propose a different mechanism to realize and manipulate spectral windings and skin modes, namely through nonconservative couplings.  
Different from coherent couplings, nonconservative couplings connect elements of the systems indirectly through intermediates with gain or loss, exhibiting non-Hermitian characteristics in the phase terms of the couplings~\cite{Yang2017,Leefmans2022,Metelmann2015,Huang2021,Huang2023}. By considering a chain of resonance modes with conservative nearest-neighbor couplings and nonconservative next-nearest-neighbor couplings, we find that diverse spectra can appear by manipulating the nonconservative coupling phases. These include closed loops with the same and opposite windings, as well as twisted windings with two oppositely oriented loops, which arise by combining the nonconservative couplings with multi-channel interference provided by the chain. 
Associated with the multi-type spectra, unipolar and bipolar NHSE with different eigenmode localizations can be observed. 
To connect with the nonreciprocity performance, we also present evidence for a close link between the nonreciprocal transmission of the chain and multi-type spectral phases as well as NHSE.

The rest of this paper is organized as follows. In Sec.~\ref{sec2}, we introduce the system model of a non-Hermitian chain with nonconservative couplings. In Sec.~\ref{sec3}, we present the spectral phase diagram, eigenstates distributions and the generalized Brillouin zone (GBZ) to illustrate the multi-type spectra and NHSE. We also show the influence of the phase and strength of the nonconservative couplings on the spectral twisting phenomenon. In Sec.~\ref{sec4}, we present evidence for a close relationship between the multi-type spectral phases and the nonreciprocal transmission of the chain. Finally, we summarize our results in Sec.~\ref{sec5}. In the Appendixes, we provide several parts of the detailed derivations, including the derivation of the effective Hamiltonian with nonconservative couplings (Appendix~\ref{appendix1}), the phase diagram of two-band winding number and Inverse Participation Ratio (IPR) of the model (Appendix~\ref{appendix2}) and the nonreciprocal energy transmission (Appendix~\ref{appendix3}).
\begin{figure}
	\centering
	\includegraphics[width=8.5cm]{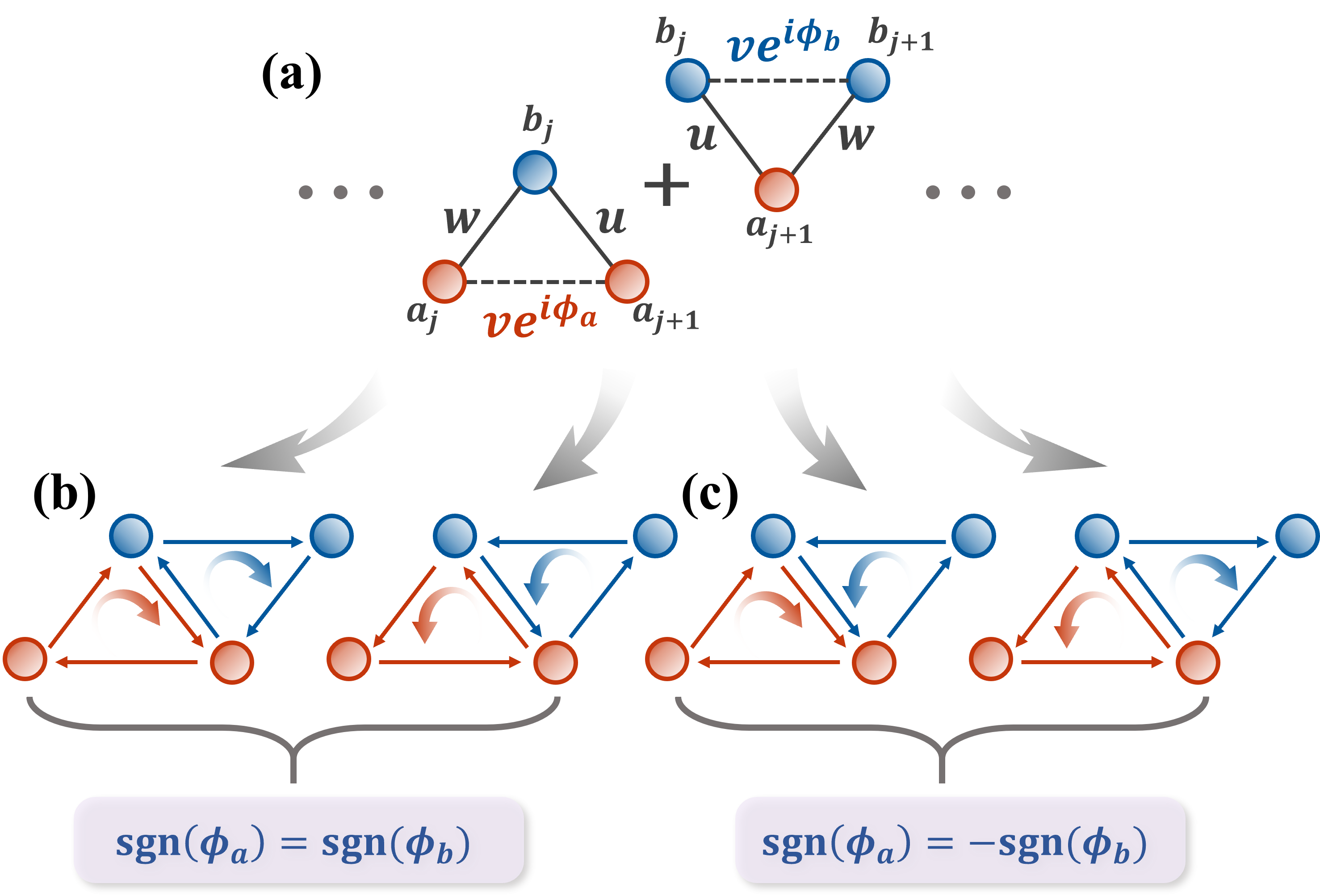}
	\caption{(a)~Schematic of the non-Hermitian chain with conservative nearest-neighbor couplings (solid lines) and nonconservative next-nearest-neighbor couplings ($ve^{i\phi_{a,b}}$, dashed lines). Red balls represent subchain $a_j$ while blue balls represent subchain $b_j$. (b-c)~Possible combinations of the upper and lower triangular plaquettes with the nonreciprocal flow. The direction of the nonreciprocal flow in each triangular plaquette (illustrated by the red and blue arrows) can be controlled by tuning the sign of the nonconservative coupling phases $\phi_{a}$ and $\phi_{b}$.}
	\label{fig1}
\end{figure}

    \section{Model}
    \label{sec2}
We consider a chain of bosonic modes with conservative nearest-neighbor couplings and nonconservative next-nearest-neighbor couplings. The system Hamiltonian is given by
\begin{equation}
	\begin{aligned}
		\hat{H} =  &\sum_{j=1}^{N} [(\delta \hat{a}_j^\dagger \hat{a}_j - \delta \hat{b}_j^\dagger \hat{b}_j)+(w\hat{a}_j^\dagger \hat{b}_j+h.c.)]\\
		&+\sum_{j=1}^{N-1} [(u\hat{b}_j^\dagger \hat{a}_{j+1}+h.c.)+v e^{i\phi_{a}}(\hat{a}_j^\dagger \hat{a}_{j+1}+h.c.)\\
		&+v e^{i\phi_{b}}(\hat{b}_j^\dagger \hat{b}_{j+1}+h.c.)],
	\end{aligned}
	\label{eqm1}
\end{equation}
where $2\delta$ is the difference of the mode resonance frequencies,
and $w$ and $u$ denote the conservative nearest-neighbor intra- and inter-hopping terms, respectively. The non-Hermiticity of the system is induced by considering nonzero phases $\phi_{a}$ and $\phi_{b}$ of the nonconservative next-nearest-neighbor couplings, where the energy flow in the coupling process $\hat{a}_j^\dagger \hat{a}_{j+1}+\hat{a}_{j+1}^\dagger \hat{a}_j$ ($\hat{b}_j^\dagger \hat{b}_{j+1}+\hat{b}_{j+1}^\dagger \hat{b}_j$) is not conserved (an example to implement the nonconservative couplings can be found in the Appendix~\ref{appendix1}). As illustrated in Fig.~\ref{fig1}(a), this periodic coupling structure can be interpreted as a combination of the upper ($a_{i}\leftrightarrow b_{i}\leftrightarrow a_{i+1}$) and lower ($b_{i}\leftrightarrow a_{i+1}\leftrightarrow b_{i+1}$) triangular plaquettes.
In each triangular plaquette, nonreciprocal energy flow can be realized by combining nonconservative couplings with two-channel interference provided by the nonzero effective flux~\cite{Huang2023}. The direction of the nonreciprocal flow in the triangular plaquette can be changed by tuning the sign of the nonconservative coupling phase $\phi_{a}$ ($\phi_{b}$). When combining the upper and lower triangular plaquettes, the nonreciprocal flow can be the same direction by choosing $\text{sgn}(\phi_{a})=\text{sgn}(\phi_{b})$, or in the opposite direction when $\text{sgn}(\phi_{a})=-\text{sgn}(\phi_{b})$, as shown in Fig.~\ref{fig1}(b) and (c).

    \section{Multi-type spectra and NHSE}
	\label{sec3}
    \subsection{Spectral phase diagram and distribution of eigenstates}
\begin{figure}
		\centering
		\includegraphics[width=8cm]{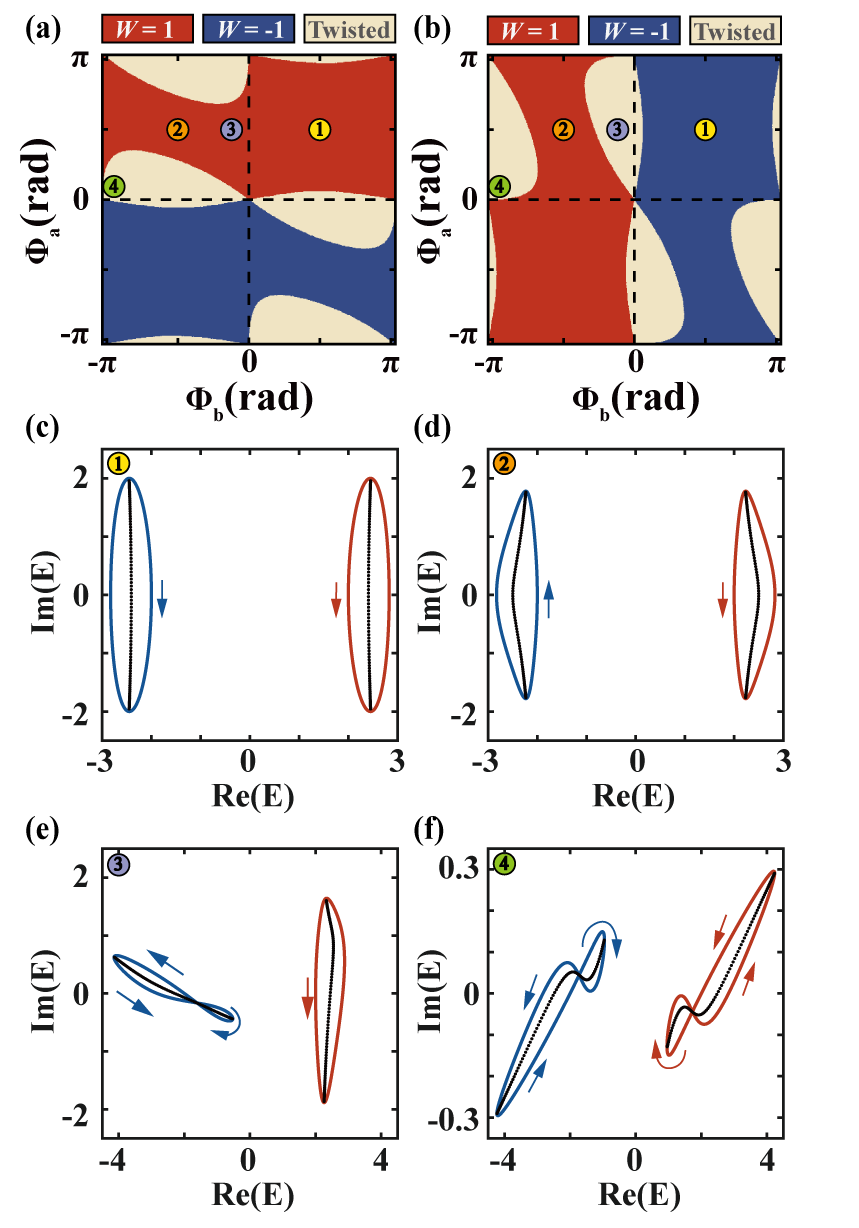}
		\caption{(a)-(b)~Phase diagram of the spectral winding number for the phase transition of two energy bands, respectively. The red region indicates the winding number $W=1$ and the blue region indicates $W=-1$. The light yellow area represents the appearance of twisted spectral winding. (c)-(f)~Spectra under PBC (red and blue for mode $a$ and $b$) and OBC (black points) of the model, with different non-conservative coupling phases $\phi_b = \pi/2,\,-\pi/2,\,-\pi/8$ and $\phi_a = \pi/2$ (c)-(e), and $\phi_b = -19\pi/20$, and $\phi_a = \pi/20$ (f), respectively. Clockwise and counterclockwise winding directions of the PBC spectrum vs the quasi-momentum $k$, as indicated by the arrows, correspond to OBC skin modes localized on the left and right, respectively. The system size is $N=100$ and other parameters are $w/\delta = 1/2,\,u/\delta = i/2,\,v/\delta = 1/2$.}
		\label{fig2}
	\end{figure}

Under the PBC, the Hamiltonian in Eq.~(\ref{eqm1}) can be diagonalized in the momentum basis with $\hat{h}(k)=\langle k| \hat{H}|k\rangle=\sum_{a_j,b_j} \langle k,a_j| \hat{H}|k,b_j\rangle|a_j\rangle\langle b_j|$, where $\hat{h}(k)$ denotes the momentum Hamiltonian. As shown in Fig.~\ref{fig2}, the energy spectra under the PBC form loops in the complex quasi-energy space, which is distinct from the two open arcs under the OBC, indicating the emergence of the NHSE in our system~\cite{Okuma2020}.
To further investigate the energy spectra and wavefunctions of our model, we characterize the spectral phases by describing the band windings~\cite{Zhang2020_wn}. For any base point $E_{b}$, the winding number of the energy spectrum is given by 
	\begin{equation}
		W=\frac{1}{2\pi i}\int_{-\pi}^{\pi}\partial_{k}\text{ln det}[\hat{h}(k)-E_{b}\mathbb{I}_{2\times2}]dk,
		\label{eqm2}
	\end{equation} 
 where $\mathbb{I}_{2\times2}$ is the two-dimensional identity matrix.
The winding number is an integer obtained by counting the number of times that the eigenenergy of $\hat{h}(k)$ wraps around the base point $E_b$ as $k$ varies from $-\pi$ to $\pi$. When the loop structure is absent in the periodic spectrum, the energy base point $E_b$ is not enclosed, resulting in $W = 0$ and the absence of the NHSE. The sign of the winding number is determined by the circling direction: anti-clockwise circling corresponds to $W = 1$ (left localization of eigenmodes), while clockwise circling corresponds to $W = -1$ (right localization of eigenmodes).

Characterizing the spectral topology by winding number $ W $, we observe that the spectral phase diagrams of the two energy bands exhibit distinct behaviors when the nonconservative coupling phases $ \phi_a $ and $ \phi_b $ are adjusted, as depicted in Fig.~\ref{fig2} (a) and (b). The red and blue regions represent the closed-loop-like band structure with winding numbers $W=1$ and $-1$~[Fig.~\ref{fig2}(c) and (d)], respectively. The light yellow area indicates the presence of the twisted winding [Fig.~\ref{fig2}(e) and (f)]. To elucidate the characteristics of spatial profiles corresponding to different spectral phases, we analyze the spatial profiles of all eigenmodes under the OBC, illustrated in Fig.~\ref{fig3}. Based on the eigenmode distributions across the lattice sites in our model, these eigenmodes are categorized into two groups: those predominantly localized on sites $ a_j $ with minimal occupation on $ b_j $, and vice versa.
In the cases where $\text{sgn}(\phi_{a}) = \text{sgn}(\phi_{b})$, the two bands typically exhibit opposite circling directions, as depicted in Fig.~\ref{fig2}. Under these conditions, the eigenmodes predominantly occupying sites $a_{j}$ localize toward the left edge, while the eigenmodes concentrated on sites $b_j$ localize at the right boundary, simultaneously. This arrangement is indicative of the NHSE, as illustrated in Fig.~\ref{fig3} (a) and (b). Conversely, when $\text{sgn}(\phi_{a}) = -\text{sgn}(\phi_{b})$, the two bands share the same winding number and circle in the same direction. This scenario results in all eigenmodes localizing at the same boundary, characteristic of the conventional unipolar NHSE, as shown in Fig.~\ref{fig3} (c)and (d).

\subsection{Twisted winding}
\begin{figure}
    \centering
    \includegraphics[width=8.5cm]{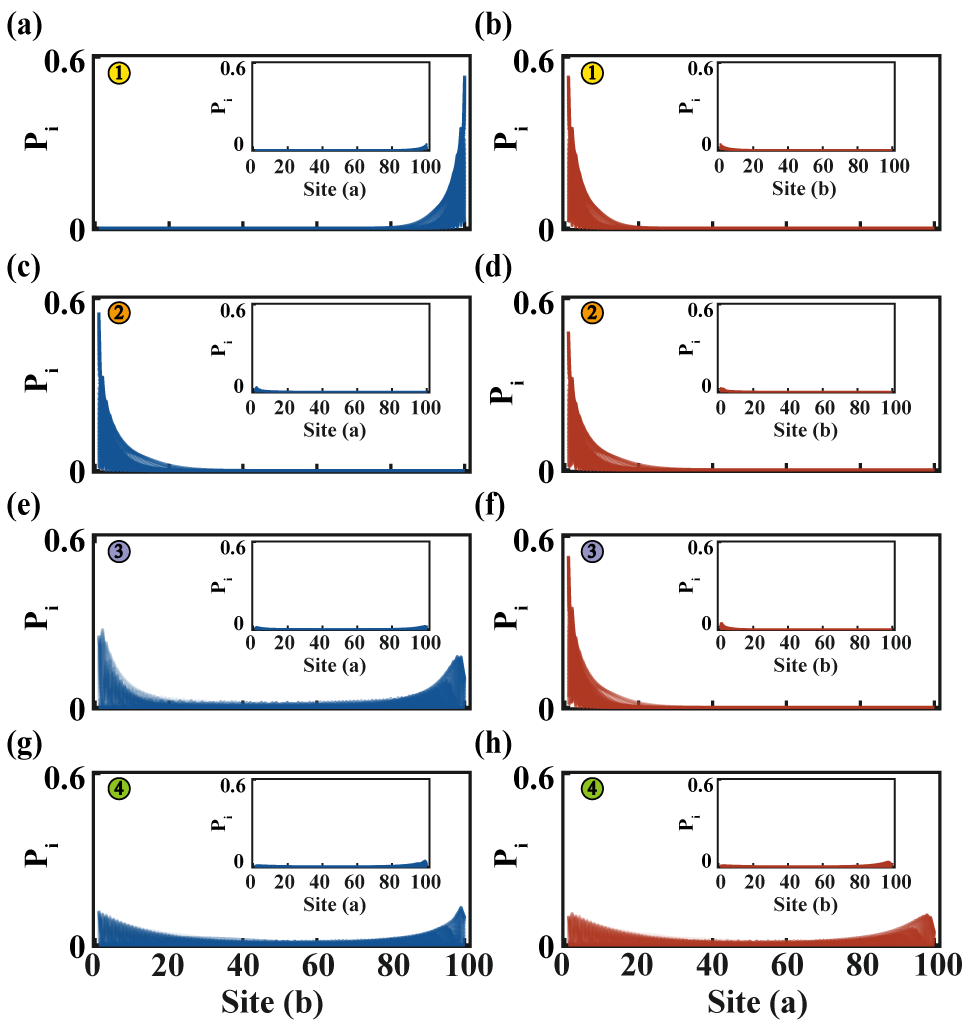}
    \caption{(a)-(h)~Distributions of all eigenmodes corresponding to phase 1-4 in Fig.~\ref{fig2} respectively. Blue (red) lines represent the distributions of the eigenstates  with eigenvalue localized in the blue (red) loop of the PBC spectrum on the $b$ ($a$) sites. Insets show the distributions of eigenmodes on the $a$ ($b$) sites, indicating a tiny population on the opposite side. The system size is $N=100$ and other parameters are $w/\delta = 1/2,\,u/\delta = i/2,\,v/\delta = 1/2$.}
    \label{fig3}
\end{figure}

Intriguingly, beyond the two regions mentioned above, there exists a unique spectrum type called twisted winding, which consists of two oppositely oriented loops in contact rather than a single loop. As illustrated in Fig.~\ref{fig2}(e), the twisted loop splits the band into two segments, with one segment circling in a clockwise direction and the other in an anticlockwise direction, resulting in two opposite winding numbers within a single band. The corresponding distributions of all eigenmodes localized at both boundaries of the system are shown in Fig.~\ref{fig3}(e). The other eigenmodes, corresponding to the spectrum with a single loop, are localized at the left boundary [Fig.~\ref{fig3}(f)]. Simultaneous twisted loops in both bands can be induced by tuning the phases $\phi_a$ and $\phi_b$, as shown in Fig.~\ref{fig2}(f). In such cases, the eigenmodes associated with the twisted loops are split, localizing towards opposite boundaries, as depicted in Fig.~\ref{fig3}(g)-(h).
Moreover, the effects of the phase difference ($\phi_{a}-\phi_{b}$) significantly influence the size of the twisting region in the phase space. Specifically, the region where $\text{sgn}(\phi_{a}) = -\text{sgn}(\phi_{b})$ is considerably larger than that in which $\text{sgn}(\phi_{a}) = \text{sgn}(\phi_{b})$.

The multi-type spectra and the NHSE localization transitions can also be observed in the context of the nonreciprocal flow in the periodic structure shown in Fig.~\ref{fig1}.
The closed-loop-like band structure with a nonzero winding number indicates the unipolar localization skin effects consistent with the unidirectional reciprocal transmission. However, twisted winding represents the respectively bidirectional localization skin effects in the corresponding subchain $a_j$ or $b_j$, which indicates competing energy transport in different directions. In our model, the main competition occurs between the inter-chain and intra-chain energy transmission channels, which may have different directions. Specifically, taking chain $a_j$ (red balls) as an example, the intra-chain energy transfer direction among modes $a_j$ may differ from the inter-chain transfer direction between $a_j$ and $b_j$ (or $a_{j+1}$ and $b_j$). When these two transfer directions are different and the inter-chain energy transfer is non-negligible, it will affect the localization direction of mode $a_j$, corresponding to the emergence of the twisted winding.
When the signs of $\phi_{a}$ and $\phi_{b}$ are the same, the opposite nonreciprocal flow from modes $a_{j}$ to $a_{j+1}$ and from $b_{j+1}$ to $b_{j}$ corresponds to the presence of the bipolar skin effects for $\text{sgn}(\phi_{a}) = \text{sgn}(\phi_{b})$. In this scenario,  the two inter-chain transmission channels between modes $a_{j+1}$ and $b_{j}$, provided by the upper and lower triangular plaquettes, interfere destructively. This results in negligible energy flow between modes $a_{j+1}$ and $b_{j}$, keeping the two bands relatively independent. Only when there is a significant difference between the phases $\phi_{a}$ and $\phi_{b}$, which results in a non-negligible effective coupling between the upper and lower triangular plaquettes, a band twist emerges.
Conversely, when the signs of $\phi_a$ and $\phi_b$ differ, the nonreciprocal flows differently, causing the same transmission from modes $b_{j}$ to $a_{j+1}$ in both the upper and lower triangular plaquettes. Constructive interference between the two inter-chain transmission channels enhances the energy flow between the modes $b_{j}$ and $a_{j+1}$, facilitating the twisting of the bands with even a small difference in the nonconservative coupling phases $\phi_{a}$ and $\phi_{b}$.

\begin{figure}
   \centering
   \includegraphics[width=8cm]{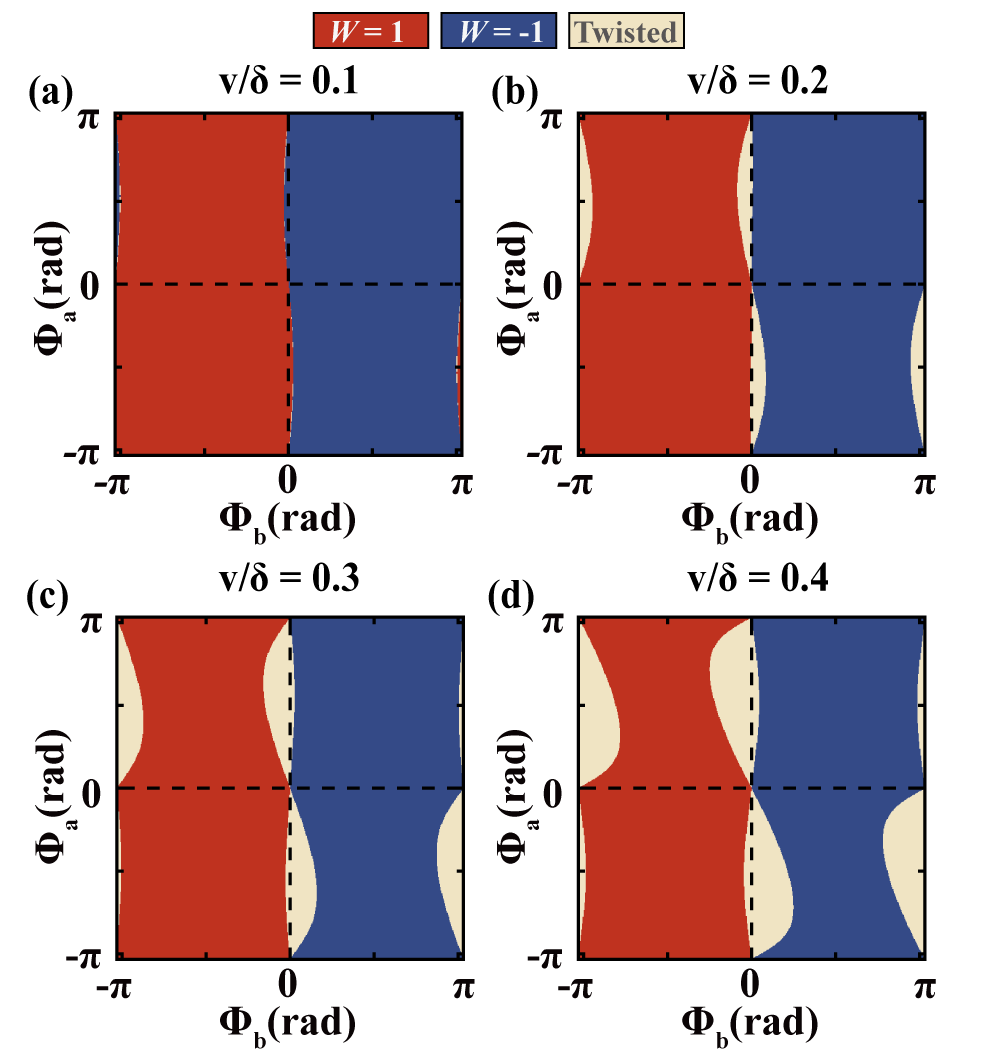}
   \caption{(a)-(d) Phase diagram of the spectral winding number for the phase transition of a single energy band with $w/\delta = 1/2,\,u/\delta = i/2$. The red region indicates the winding number $W=1$ and the blue region indicates $W =-1$. The light yellow area represents the appearance of twisted spectral winding. The strength of the nonconservative coupling $v/\delta$ increases from 0.1 to 0.4.}
   \label{fig4}
\end{figure}

 Moreover, we investigate the influence of the nonconservative coupling strength on the phase diagram of the spectral winding number. In Fig.~\ref{fig4}, we present the phase diagram of the spectral winding number for the phase transition of a single energy band. The red region indicates the winging number $W=1$ and the blue region indicates $W=-1$. The light yellow area represents the appearance of twisted spectral winding. As the nonconservative coupling strength $v/\delta$ increases from 0.1 to 0.4, we observe an expansion of the twisting region in the phase diagram. This expansion can be attributed to the increasing competition between the inter-chain and intra-chain energy transmission in the system.
    
\subsection{GBZ}
    \begin{figure}
   	\centering
   	\includegraphics[width=7.9cm]{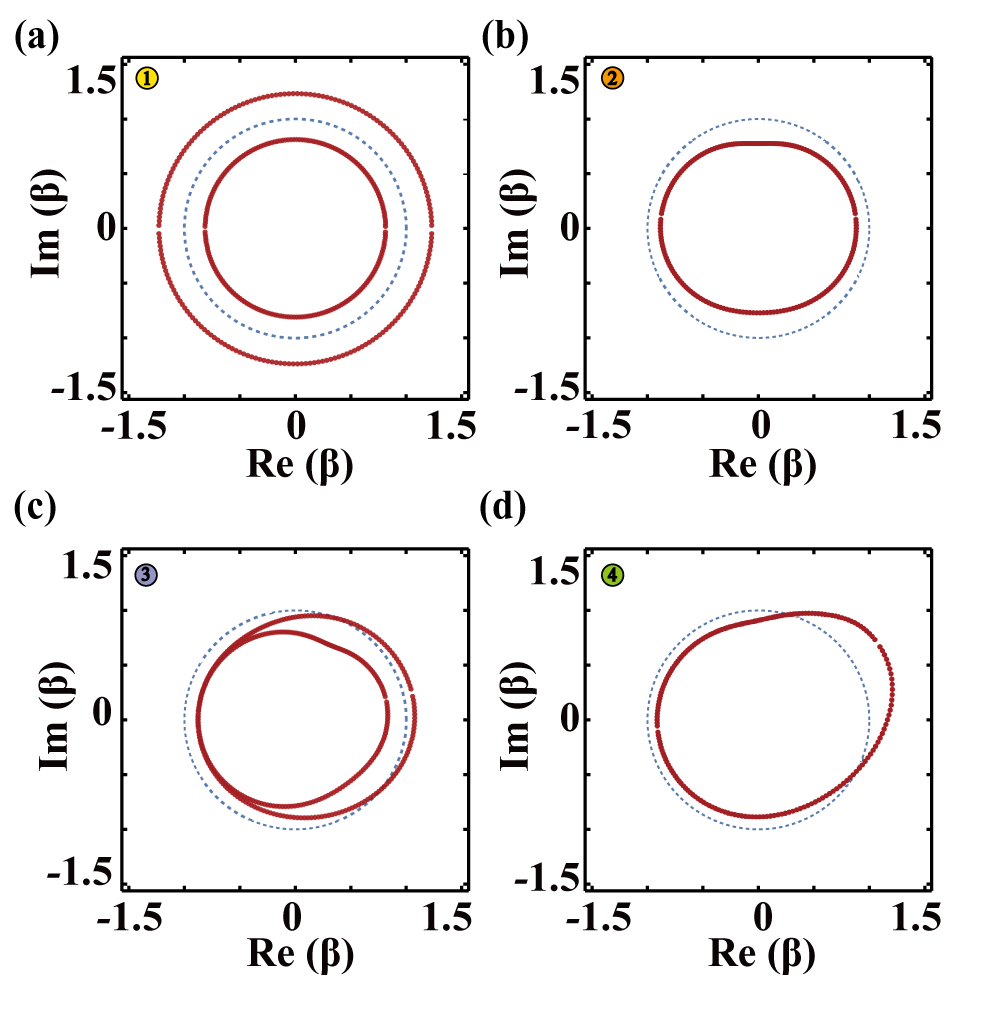}
   	\caption{(a)-(d) The Generalized Brillouin zone (red dots) of the system with different nonconservative coupling phases $\phi_b\ =\ \pi/2$,$-\pi/2,-\pi/6$ and $\phi_a\ =\ \pi/2$ (a)-(c), and $\phi_b\ =\ -\pi/6$, $\phi_a\ =\ 5\pi/6 $ (d), respectively. The blue dash line is the unit circle. The system size is $N=100$ and other parameters are $w/\delta = 1/2,\,u/\delta = i/2,\,v/\delta = 1/2$.}
   	\label{fig5}
   \end{figure}
    
    The properties of the non-Hermitian skin effect can also be characterized by the GBZ~\cite{Yao2018}, which is calculated by
    \begin{equation}
        f(\beta,E) = \text{det}[E\cdot\mathbb{I}_{2\times2} - h(\beta)] = 0,
    \end{equation}
    with $h(\beta) = h(e^{ik}\rightarrow \beta)$ and $E$ being the eigenvalues under the OBC. Here, $f(\beta, E)$ is a quartic equation for $\beta$, having four solutions $\beta_i$, with $|\beta_1| \leqslant |\beta_2| \leqslant |\beta_3| \leqslant |\beta_4|$. The trajectory of solution $\beta$ satisfying $|\beta_2|=|\beta_3|$ gives rise to the GBZ. 

    Fig.~\ref{fig5} shows the GBZs for the different nonconservative coupling phases. When the nonconservative coupling phases $\phi_b\ =\ \pi/2$ and $\phi_a = \pi/2$, the spectra winding numbers of the two bands are $W_1=1$ and $W_2=-1$. As seen in Fig.~\ref{fig5}(a), one set of sub-GBZs with $\text{Re}(E) > 0$ is located inside the unit circle (blue dashed lines), indicating localization on the left boundary. However, another set of sub-GBZs with $\text{Re}(E) < 0$ lies outside the unit circle, indicating localization on the right side. When $\phi_b\ =\ -\pi/2$ and $\phi_a = \pi/2$, all spectra winding numbers are $W_1 = W_2 = 1$, indicating that all skin modes are localized at the left boundary. The corresponding GBZs are located inside the unit circle, as shown in Fig.~\ref{fig5}(b). In the twisted winding region [Fig.~\ref{fig5} (c) and (d)], GBZs intersect with the unit circle. These intersection points are Bloch points, indicating that the corresponding eigenstates of the system exhibit Bloch-wave-like behavior.

	\section{Connection with nonreciprocity}
	\label{sec4}
Extending the analysis of the connection between the nonreciprocity performance of the chain and the NHSE shown in different phase regions, we consider two different nonreciprocal transmission cases by choosing the leftmost mode $a_{1}$ ($b_{1}$) and rightmost mode $a_{N}$ ($b_{N}$) to connect the input and output field, respectively. 
Therefore, the nonreciprocity ratios of the two cases can be defined as $I_a  =T_{a\leftarrow}/T_{a\rightarrow}=|S_{1,2N-1}/S_{2N-1,1}|^2$, $I_b =T_{b\leftarrow}/T_{b\rightarrow}= |S_{2,2N}/S_{2N,2}|^2$, 
which can be calculated by solving the system scattering matrix $S$ numerically (see the Appendix~\ref{appendix3} for more details). Once the off-diagonal elements of $S$ satisfy $|S_{1,2N-1}|\neq |S_{2N-1,1}|$ ($|S_{2,2N}|\neq |S_{2N,2}|$), the nonreciprocal transmission in the chain can be achieved. When the transmission efficiency for the backward (forward) direction is larger than that for the forward (backward) direction for the chain, i.e., $I_{a/b}>1$ ($I_{a/b}<1$), the nonreciprocity direction of the chain is backward (forward).

\begin{figure}
		\centering
		\includegraphics[width=8.2cm]{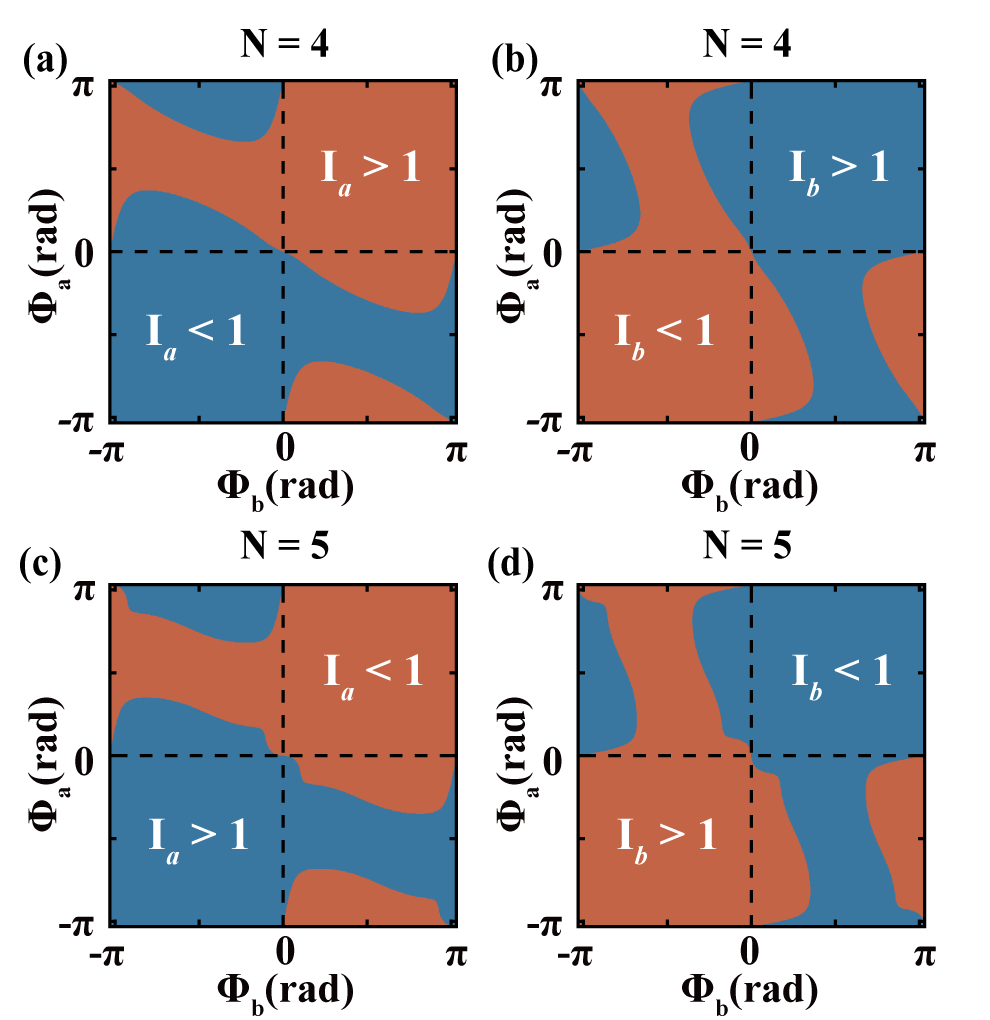}
		\caption{Diagrams of nonreciprocity ratios as functions of the nonconservative coupling phases $\phi_{a}$ and $\phi_{b}$, with the input and output field connected to mode $a_1/a_N$(a)(c) and $b_1/b_N$(b)(d) with different sizes of the chain. The red indicates the nonreciprocity ratio $I_{a/b}>1$, and the blue region indicates $I_{a/b}<1$. The size of the chain is $N=4$(a)(b) and $5$(c)(d) and  the other parameters are $w/\delta = 1/2,\,u/\delta = i/2,\,v/\delta = 1/2$.} 
	\label{fig6}
    \end{figure}

	The nonreciprocity ratios of the two different settings for the input and output field connecting with the mode $a_{1}$/$a_{N}$ and $b_{1}$/$b_{N}$ are illustrated in Fig.~\ref{fig6} (a)(c) and (b)(d), respectively. The red (blue) region represents the nonreciprocity ratio  $I_{a/b} > 1$ ($I_{a/b} < 1$), indicating the backward (forward) nonreciprocity direction for the chain. Comparing the spectral phase diagram shown in Fig.~\ref{fig2} (a) and (b), we can find the nonreciprocal transmission shows similar behaviors with the eigenmode localization of NHSE. When tuning the phases as $\text{sgn}(\phi_{a}) = -\text{sgn}(\phi_{b})<0$ and $|\phi_a|$ and $|\phi_b|$ are around $\pi/2$, both $I_{a}$ and $I_{b}$ are greater than 1 for left eigenmode localization and less than 1 for right eigenmode localization of unipolar NHSE. However, the nonreciprocity directions of the two input and output settings will become opposite when tuning the nonconservative coupling phases to the region of closed loops with opposite windings. We can find that the nonreciprocity ratio $I_{a}>1$ while $I_{b}<1$ when $\text{sgn}(\phi_{a}) = \text{sgn}(\phi_{b})>0$, and $I_{a}<1$ while $I_{b}>1$ when $\text{sgn}(\phi_{a}) = \text{sgn}(\phi_{b})<0$ [Fig.~\ref{fig6} (a)(c) and (b)(d)]. 
Therefore, the opposite nonreciprocity directions of the two settings can connect with the opposite localization of the two groups of eigenmodes, indicating a close link between the bipolar NHSE and nonreciprocity. 

Moreover, as we approach the band twisting region, the change of the nonreciprocity direction is determined by the values of the nonconservative coupling phases and the chain size $N$ together. As illustrated in Fig.~\ref{fig6}, comparing the Fig.~\ref{fig6}(a)(b) with (c)(d), which represent different system sizes, we observe that the change in nonreciprocity direction is influenced by both the nonconservative coupling phases and the chain size $N$. Due to the rapid decrease in transmission efficiency with increasing $N$, we limit our analysis of nonreciprocity to chains where $N\leq 5$. 
    \section{Conclusion}
    \label{sec5}
In conclusion, we elucidated the manipulation of the NHSE through nonconservative couplings that extend beyond nearest-neighbor couplings, demonstrating multi-type spectral phases and their impact on mode localizations. By tuning the phases of the nonconservative couplings, we achieved diverse spectral structures associated with both unipolar and bipolar NHSE localizations. Moreover, we have evidenced that multi-type spectral phases as well as NHSE are closely linked to the nonreciprocal transmission of the chain.
Our approach is also experimentally feasible as nonconservative couplings have been realized in various systems, including optical systems~\cite{Arwas2022,Bergman2021}, room-temperature atomic ensembles~\cite{Peng2016} and magnonics system~\cite{Yang2020,Zou2024}. 
Our findings not only provide new insights into the control of NHSE but also pave the way for advanced applications in controlling nonreciprocal energy flow.

    \begin{acknowledgements}
		This work is supported by the Beijing Natural Science Foundation (Grant No. Z240007), National Natural Sci- ence Foundation of China (Grants No. 12474354 and No. 12125402), and the Innovation Program for Quantum Science and Technology (Grant No. 2021ZD0301500), also the Fundamental Research Funds for the Central Universities.
    \end{acknowledgements}

    \begin{appendix}
        \section{Derivation of the effective Hamiltonian with nonconservative couplings}
   \label{appendix1}
  In this section, we present an example to implement the nonconservative couplings~\cite{Huang2021,Huang2023}. Considering an array of bosonic modes $a_j$ and $b_j$ ($j = 1,\,2,...,N$), which do not directly interact with each other. Interactions between $a_j$ and $a_{j+1}$ are mediated by an auxiliary mode $q_j$ subject to gain or loss~\cite{Yang2017,Leefmans2022}. Similarly, mode $b_j$ interacts with $b_{j+1}$ through auxiliary mode $o_j$. In the reference frame rotating at frequency $\omega_0$, we apply the transformations $\hat{a}_j \to \hat{a}_je^{-i\omega_0 t}, \hat{b}_j \to \hat{b}_je^{-i\omega_0 t}, \hat{q}_j\to \hat{q}_je^{-i\omega_0 t}$ and $\hat{o}_j \to \hat{o}_je^{-i\omega_0 t}$. Therefore, the system Hamiltonian reads ($\hbar = 1$)
  \begin{widetext}
   \begin{equation}
   	\begin{aligned}
   		H_{\text{o}}= &\sum_{j=1}^{N} [-\Delta_a \hat{a}_j^\dagger \hat{a}_j-\Delta_b \hat{b}_j^\dagger \hat{b}_j - (\Delta_o+i\frac{\gamma_j}{2}) \hat{o}_j^\dagger \hat{o}_j - (\Delta_q+i\frac{\kappa_j}{2}) \hat{q}_j^\dagger \hat{q}_j +(w\hat{a}_j^\dagger \hat{b}_j+w^*\hat{b}_j^\dagger \hat{a}_j)] \\
   		& +\sum_{j=1}^{N-1} [( g_{a,L} \hat{a}_j^\dagger +g_{a,R}\hat{a}_{j+1}^\dagger)  \hat{q}_j +( g_{a,L}^*  \hat{a}_j +g_{a,R}^* \hat{a}_{j+1})  \hat{q}_j^\dagger +( g_{b,L}  \hat{b}_j^\dagger +g_{b,R} \hat{b}_{j+1}^\dagger)  \hat{o}_j \\&+( g_{b,L}^*  \hat{b}_j +g_{b,R}^*  \hat{b}_{j+1})  \hat{o}_j^\dagger + (u\hat{b}_j^\dagger \hat{a}_{j+1}+u^*\hat{a}_{j+1}^\dagger \hat{b}_j)].
   	\end{aligned}
   	\label{eq1}
   \end{equation}
   \end{widetext}
   Here, $\Delta_{a,b,q,o} = \omega_0 - \omega_{a,b,q,o}$ are the detunings of the corresponding modes. $\gamma_j$ and $\kappa_j$ represent the energy decay rates of modes $q_j$ and $o_j$, respectively. The operators $\hat{a}_j^\dagger$, $\hat{b}_j^\dagger$, $\hat{q}_j^\dagger$ and $\hat{o}_j^\dagger$, along with $\hat{a}_j$, $\hat{b}_j$, $\hat{q}_j$ and $\hat{o}_j$, represent the creation and annihilation operators for modes $a_j$, $b_j$, $q_j$ and $o_j$, respectively, and obey bosonic commutation relations. The coupling coefficients are $g_{a,R}$ between modes $a_{j+1}$ and $q_j$, $g_{a,L}$ between modes $a_j$ and $q_j$, $g_{b,R}$ between modes $b_{j+1}$ and $o_j$, and $g_{b,L}$ between modes $b_j$ and $o_j$, respectively, with subscripts "$L$" for left and "$R$" for right. 
   
   Since modes $q_j$ and $o_j$ serve only the purpose of mediating interactions, we only consider the noise fluctuations of modes $q_j$ and $o_j$ which are connected to the input and output fields. The Langevin equations can be written as
   \begin{widetext}
    \begin{equation}
   	\begin{aligned}
   		\frac{d\hat{a}_j}{dt} & = i\Delta_a \hat{a}_j - i(w\hat{b}_j + u^*\hat{b}_{j-1}^\dagger + g_{a,L}\hat{q}_j + g_{a,R}\hat{q}_{j-1}),\\
   		\frac{d\hat{b}_j}{dt} & = i\Delta_b \hat{b}_j - i(w^*\hat{a}_j + u\hat{a}_{j+1}^\dagger+ g_{b,L}\hat{o}_j + g_{b,R}\hat{o}_{j-1}),\\
   		\frac{d\hat{q}_j}{dt} & =  (i\Delta_q - \frac{\gamma_j}{2}) \hat{q}_j - i (g_{a,L}^*\hat{a}_j + g_{a,R}^*\hat{a}_{j+1}) + \sqrt{\gamma_{j}^{ex}}\hat{q}_j^{\text{in}} + \sqrt{\gamma_j^0}\hat{f}_j^{\text{in}},\\
   		\frac{d\hat{o}_j}{dt} & = (i\Delta_o - \frac{\kappa_j}{2}) \hat{o}_j - i(g_{b,L}^*\hat{b}_j + g_{b,R}^*\hat{b}_{j+1})+ \sqrt{\kappa_{j}^{ex}}\hat{o}_j^{\text{in}}  + \sqrt{\kappa_j^0}\hat{f}_j^{\text{in}}.
   	\end{aligned}
   	\label{eq2}
   \end{equation}
   \end{widetext}
 
   The damping rates associated with the input fields $q_j^{\text{in}}$ and $o_j^{\text{in}}$ are denoted by $\gamma_{j}^{\text{ex}}$ and $\kappa_{j}^{\text{ex}}$, respectively. The intrinsic damping rates are $\gamma_j^0$ and $\kappa_j^0$, with the associated noise given by $f_j^{\text{in}}$. Assuming that the energy decay rates of modes $q_j$ and $o_j$ are identical and the intrinsic rates are much smaller than the damping rates, we have $\gamma_j=\gamma_j^{\text{ex}}+\gamma_j^0\approx\gamma_{j}^{\text{ex}}=\gamma$ and $\kappa_j=\kappa_j^{\text{ex}}+\kappa_j^0\approx\kappa_{j}^{\text{ex}}=\kappa$. Additionally, we consider the vacuum input field, i.e., $\langle \hat{q}_j^{\text{in}\dagger} \hat{q}_j^{\text{in}}\rangle=\langle \hat{o}_j^{\text{in}\dagger} \hat{o}_j^{\text{in}} \rangle = 0$. When  the detunings $\Delta_{q}$ and $\Delta_{o}$ or the energy decay rates $\gamma$ and $\kappa$ are much larger than the coupling coefficients, i.e., $|\Delta_q+i\frac{\gamma}{2}|\gg (|g_{a, L}|,|g_{a, R}|)$ and $|\Delta_o+i\frac{\kappa}{2}|\gg (|g_{b, L}|,|g_{b, R}|)$, the connecting modes $q_i$ and $o_i$ can be adiabatically eliminated as
   \begin{equation}
   	\begin{aligned}
   		\hat{q}_j & = \frac{g_{a,L}^*\hat{a}_j + g_{a,R}^*\hat{a}_{j+1}}{\Delta_q+i\gamma/2},\\
   		\hat{o}_j & = \frac{g_{b,L}^*\hat{b}_j + g_{b,R}^*\hat{b}_{j+1}}{\Delta_o+i\kappa/2}.
   	\end{aligned}
   	\label{eq3}
   \end{equation}
    For convenience, we assume that $g_{a,L}, g_{a,R}, g_{b,L}$ and $ g_{b,R}$ are real numbers ($\in \mathbb{R}$). By substituting Eq.~(\ref{eq3}) into Eq.~(\ref{eq2}), we obtain
    \begin{widetext}
   \begin{equation}
   	\begin{aligned}
   		\frac{d\hat{a}_j}{dt} & = (i\Delta_a - i\frac{|g_{a,L}|^2+|g_{a,R}|^2}{\Delta_q+i\gamma/2}) \hat{a}_j - i( \frac{g_{a,L}g_{a,R} \hat{a}_{j+1} +g_{a,L}g_{a,R} \hat{a}_{j-1}}{\Delta_q+i\gamma/2}+ w\hat{b}_j + u^*\hat{b}_{j-1})\\
   		& = (i\Delta_a +i\Omega_a) \hat{a}_j - i(h_{j,j+1} \hat{a}_{j+1} +h_{j,j-1} \hat{a}_{j-1}+ w\hat{b}_j + u^*\hat{b}_{j-1}),\\
   		\frac{d\hat{b}_j}{dt} & = (i\Delta_b -i\frac{|g_{b,L}|^2+|g_{b,R}|^2}{\Delta_o+i\kappa/2}) \hat{b}_j - i(\frac{g_{b,L}g_{b,R} \hat{b}_{j+1} +g_{b,L}g_{b,R} \hat{b}_{j-1}}{\Delta_o+i\kappa/2}+ w^*\hat{a}_j + u\hat{a}_{j+1}),\\
   		& = (i\Delta_b +i\Omega_b) \hat{b}_j - i(t_{j,j+1} \hat{b}_{j+1} +t_{j,j-1} \hat{b}_{j-1}+ w^*\hat{a}_j + u\hat{a}_{j+1}),
   	\end{aligned}
   	\label{eq4}
   \end{equation}
    \end{widetext}
    where 
        \begin{equation}
           	\begin{aligned}
           		&h_{j,j+1} = h_{j,j-1}  = \frac{g_{a,L}g_{a,R}}{\Delta_q+i\gamma/2},\\
           		&t_{j,j+1} = t_{j,j-1}  = \frac{g_{b,L}g_{b,R}}{\Delta_o+i\kappa/2},\\
                    &\Omega_a  =  -\frac{|g_{a,L}|^2+|g_{a,R}|^2}{\Delta_q+i\gamma/2},\\
   		      &\Omega_b  =  -\frac{|g_{b,L}|^2+|g_{b,R}|^2}{\Delta_o+i\kappa/2}.
           	\end{aligned}
           	\label{eq5}
           \end{equation}
    $h_{j,j+1}$ represents the backward coupling from $a_{j+1}$ to $a_j$ while $h_{j+1,j}$ represents the forward coupling from $a_{j}$ to $a_{j+1}$. Similarly, $t_{j,j+1}$ represents the backward coupling from $b_{j+1}$ to $b_{j}$, and $t_{j+1,j}$ represents the forward coupling from $b_{j}$ to $b_{j+1}$. $\Omega_a$ and $\Omega_b$ describe the resonance shift and broadening for modes $a_j$ and $b_j$ induced by the intermediate modes $q_j$ and $o_j$, respectively.

    \begin{figure}
   	\centering
   	\includegraphics[width=8cm]{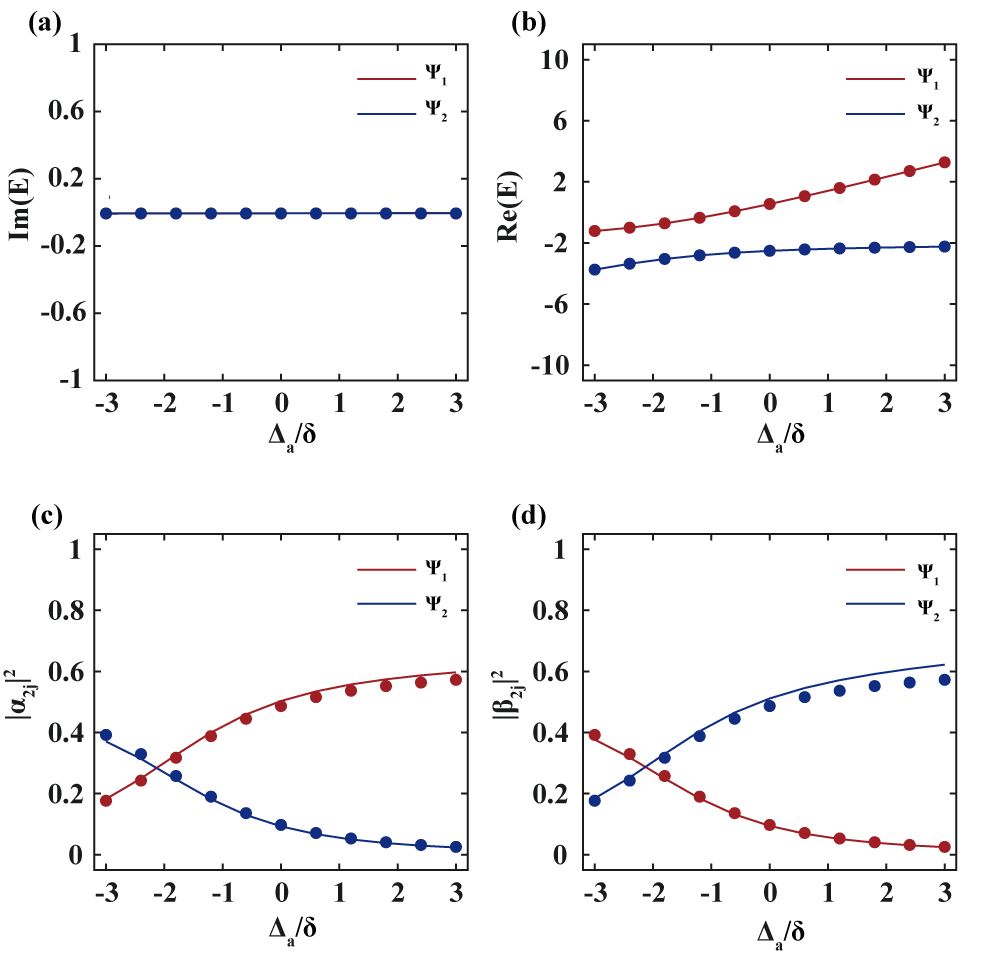}
   	\caption{Imaginary (a) and real (b) parts of the eigenenergy of the original Hamiltonian [Eq.~\ref{eq1}] (curves) and the effective Hamiltonian [Eq.~\ref{eq8}] (dots) as functions of $\Delta_a$. Occupation of modes $a_2$(c) and $b_2$(d) of the original Hamiltonian (curves) and the effective Hamiltonian (dots) as functions of $\Delta_a$. The dots and curves indicate the distributions denoted by the eigenmodes $\Psi_1$ (red) and $\Psi_2$ (blue). Other parameters are $\gamma/\delta = 5,\,\kappa = \gamma,\,\Delta_a = 0.4\gamma,\,\Delta_b = -0.4\gamma,\,\Delta_o = \Delta_q = 1.5\gamma,\,g_a = 0.1\sqrt{\gamma*|\Delta_o/\gamma+i/2|},\, g_b = 0.1\sqrt{\kappa*|\Delta_q/\kappa+i/2|},\,w = \gamma/10,\,u = 3i/\gamma$ and the number of subcells is $N = 2$.}
   	\label{figs1}
   \end{figure}
   
   Specifically, we consider the amplitudes of the effective couplings to satisfy $g_{a,L}=g_{a,R}=g_a$ and $g_{b,L}=g_{b,R}=g_b$. Therefore, Eq.~(\ref{eq5}) can be expressed as
   \begin{equation}
       \begin{aligned}
           		&h_{j,j+1} = h_{j,j-1} = v_a e^{i\phi_a}  = \frac{g_a^2}{\Delta_q+i\gamma/2},\\
           		&t_{j,j+1} = t_{j,j-1} =  v_b e^{i\phi_b} = \frac{g_b^2}{\Delta_o+i\kappa/2},\\
                    &\Omega_a  =  -\frac{2g_a^2}{\Delta_q+i\gamma/2}=-2v_ae^{i\phi_a},\\
   		      &\Omega_b  =  -\frac{2g_b^2}{\Delta_o+i\kappa/2}=-2v_be^{i\phi_b}.
           	\end{aligned}
   	\label{eq6}
   \end{equation}
   The amplitudes and phases of the nonconservative couplings are
   \begin{equation}
   	\begin{aligned}
   		& v_a  = \frac{g_a^2}{\sqrt{\Delta_q^2+\gamma^2/4}},\\
            & v_b  = \frac{g_b^2}{\sqrt{\Delta_o^2+\kappa^2/4}},\\
   		& \phi_a = -arg(\Delta_q+i\frac{\gamma}{2}),\\
   		& \phi_b = -arg(\Delta_o+i\frac{\kappa}{2}).
   	\end{aligned}
   	\label{eq7}
   \end{equation}
   Where $v_a$ and $v_b$ are the amplitudes of the nonconservative couplings, and $\phi_a$ and $\phi_b$ are the phases of the nonconservative couplings.

    Assuming the amplitudes of the nonconservative coupling are the same ($v_a = v_b =v$), we can also derive the effective non-Hermitian Hamiltonian from Eq.~(\ref{eq4}), which is given as
    \begin{widetext}
   \begin{equation}
   	\begin{aligned}
   		H_{\text{eff}}  = & H_{\text{eff}}^{\text{intra}} +H_{\text{eff}}^{\text{inter}},\\
   		H_{\text{eff}}^{\text{intra}} = & \sum_{j=1}^{N} [-\Delta_a\hat{a}_j^\dagger \hat{a}_j  -\Delta_b\hat{b}_j^\dagger \hat{b}_j + (w\hat{a}_j^\dagger \hat{b}_j+w^*\hat{b}_j^\dagger \hat{a}_j)+2v(e^{i\phi_a}\hat{a}_j^\dagger \hat{a}_j + e^{i\phi_b}\hat{b}_j^\dagger \hat{b}_j )],\\
   		H_{\text{eff}}^{\text{inter}} = & \sum_{j=1}^{N-1} [v e^{i\phi_a}(\hat{a}_j^\dagger \hat{a}_{j+1}+\hat{a}_{j+1}^\dagger \hat{a}_j) +v e^{i\phi_b}(\hat{b}_j^\dagger \hat{b}_{j+1}+\hat{b}_{j+1}^\dagger \hat{b}_j)+ (u\hat{b}_j^\dagger \hat{a}_{j+1}+u^*\hat{a}_{j+1}^\dagger \hat{b}_j)].
   	\end{aligned}
   	\label{eq8}
   \end{equation}     
    \end{widetext}
   The eigenmode $\Psi_j$ can be defined as $\Psi_j = \alpha_{1j} a_1 + \alpha_{2j} a_2 + \beta_{1j} b_1 + \beta_{2j} b_2$ ($j = 1, 2, 3, 4$) when $N=2$. In Fig.~\ref{figs1}, we plot the imaginary and real parts of the eigenenergy, along with the occupation $|\alpha_{2j}|^2$ and $|\beta_{2j}|^2$ of modes $a_2$ and $b_2$, as functions of the detuning $\Delta_a$. Notably, the results obtained by diagonalizing the effective Hamiltonian Eq.~(\ref{eq8}) match well with those obtained from the original Hamiltonian Eq.~(\ref{eq1}), as shown in Fig.~\ref{figs1}.

   Note that the last term of $H_{\text{eff}}^{\text{intra}}$ describes on-site energy nonconservation, which causes a global energy shift of the system. To focus on the non-Hermiticity induced by the nonconservative couplings, we ignore this term in the Hamiltonian. Therefore, Eq.~(\ref{eq8}) can be expressed as 
   \begin{widetext}
   \begin{equation}
   	\begin{aligned}
   		H_\text{r} =  &\sum_{j=1}^{N} [(\delta_a \hat{a}_j^\dagger \hat{a}_j + \delta_b \hat{b}_j^\dagger \hat{b}_j)+(w\hat{a}_j^\dagger \hat{b}_j+w^*\hat{b}_j^\dagger \hat{a}_j)]\\
   		&+\sum_{j=1}^{N-1} [v e^{i\phi_{a}}(\hat{a}_j^\dagger \hat{a}_{j+1}+\hat{a}_{j+1}^\dagger \hat{a}_j) +v e^{i\phi_{b}}(\hat{b}_j^\dagger \hat{b}_{j+1}+\hat{b}_{j+1}^\dagger \hat{b}_j)+(u\hat{b}_j^\dagger \hat{a}_{j+1}+u^*\hat{a}_{j+1}^\dagger \hat{b}_j)],
   	\end{aligned}
   	\label{eq9}
   \end{equation}
   \end{widetext}
   where $\delta_a = - \Delta_a\in\mathbb{R}$ and $\delta_b = - \Delta_b\in\mathbb{R}$.
   Considering the unitary transformation $U(t) = \text{exp}[-i(\frac{\delta_a+\delta_b}{2}\hat{a}_j^\dagger\hat{a}_j + \frac{\delta_a+\delta_b}{2}\hat{b}_j^\dagger\hat{b}_j)t]$, we have $a_j \rightarrow a_je^{-i\frac{\delta_a+\delta_b}{2}t}$ and $b_j \rightarrow b_je^{-i\frac{\delta_a+\delta_b}{2}t}$. After the transformation, the Hamiltonian becomes
   \begin{widetext}
   \begin{equation}
       \begin{aligned}
           \tilde{H}_\text{r} & = U(t) H_\text{r} U(t)^\dagger - iU(t)(\partial_t U(t)^\dagger) \\
           & = \sum_{j=1}^{N} [(\frac{\delta_a-\delta_b}{2} \hat{a}_j^\dagger \hat{a}_j - \frac{\delta_a-\delta_b}{2} \hat{b}_j^\dagger \hat{b}_j)+(w\hat{a}_j^\dagger \hat{b}_j+w^*\hat{b}_j^\dagger \hat{a}_j)]\\
           &+\sum_{j=1}^{N-1} [v e^{i\phi_{a}}(\hat{a}_j^\dagger \hat{a}_{j+1}+\hat{a}_{j+1}^\dagger \hat{a}_j) +v e^{i\phi_{b}}(\hat{b}_j^\dagger \hat{b}_{j+1}+\hat{b}_{j+1}^\dagger \hat{b}_j)+(u\hat{b}_j^\dagger \hat{a}_{j+1}+u^*\hat{a}_{j+1}^\dagger \hat{b}_j)],
       \end{aligned}
   \end{equation}     
   \end{widetext}
   Therefore, for simplicity, we can consider the case where $\delta_a = -\delta_b = \delta$. 
   The manifestation of NHSE in the system is primarily determined by the relative magnitudes of $\delta$, $u$, $w$, and $v$. Modifying the magnitude of $\delta$ while maintaining constant coupling strengths is equivalent to adjusting the coupling strengths while keeping $\delta$ fixed. Moreover, the non-Hermiticity is only induced by the nonconservative coupling we focus on. Therefore, the Hamiltonian becomes Eq.~(1) in our main text by defining $(\delta_a-\delta_b)/2 \equiv \delta$. In the following calculation, we choose $\delta$ as the unit value.
   Applying the Fourier transformation $\hat{a}_{j}=\frac{1}{\sqrt{N}}\sum\nolimits_{k}\hat{a}_{k}e^{ikj}$, $\hat{a}_{j}^{\dagger}=\frac{1}{\sqrt{N}}\sum\nolimits_{k}\hat{a}_{k}^{\dagger}e^{-ikj}$, $\hat{b}_{j}=\frac{1}{\sqrt{N}}\sum\nolimits_{k}\hat{b}_{k}e^{ikj}$ and $\hat{b}_{j}^{\dagger}=\frac{1}{\sqrt{N}}\sum\nolimits_{k}\hat{b}_{k}^{\dagger}e^{-ikj}$, where $k=2\pi i/N$ with $i=0,1,..., N-1$, and using the formulas $\frac{1}{N}\sum_{k}e^{i(j-j')k}=\delta_{jj'}$ and $\frac{1}{N}\sum_{j}e^{i(k-k')j}=\delta_{kk'}$, we can transform the Hamiltonian Eq.~\eqref{eq9} from real space to momentum space as
   \begin{equation}
   	H_\text{k}=
   	\begin{pmatrix}
   		\delta +2\,ve^{i\phi_a}\cos(k) & w+u^* e^{-ik} \\
   		w^*+u e^{ik} & -\delta + 2\,ve^{i\phi_b}\cos(k)
   	\end{pmatrix}.
   	\label{eq10}
   \end{equation}
   When $\phi_a = \phi_b = 0$, the non-Hermitian momentum space Hamiltonian $H_k$ degenerates into a Hermitian one. 

   \section{Phase diagram of two-band winding number and IPR}
   \label{appendix2}
      \begin{figure}
   	\centering
   	\includegraphics[width=8cm]{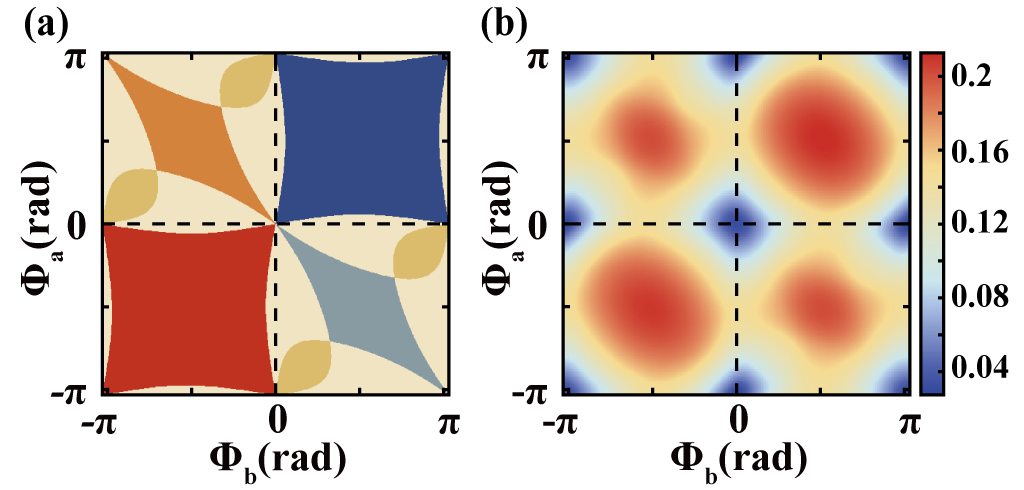}
   	\caption{(a) Phase diagram of the spectral winding number for the phase transition of two energy bands. The orange region indicates that the winding numbers of two bands are $W_1 = W_2 =1$ and the light blue region indicates $W_1 = W_2 = -1$. Dark blue and red region represent $W_1 = 1, W_2 = -1$ and $W_1 = -1, W_2 = 1$, respectively. The light yellow region represents the appearance of twist topology in one band while the dark yellow region represents the simultaneous twisted loops in both bands. Other parameters are $w/\delta = 1/2,\,u/\delta = i/2,\,v/\delta = 1/2$. (b) Diagram of the IPR of all eigenmodes. The color bar indicates the sIPR of the eigenmodes.}
   	\label{figs6}
   \end{figure}
    As shown in Fig.~\ref{figs6}(a), the orange and light blue regions represent two closed-loop-like bands encircling in the same direction with the same winding numbers $W_1 = W_2 = 1$ and $-1$, respectively, indicating the appearance of unipolar NHSE. Dark blue and red regions represent two closed-loop-like bands encircling in opposite directions with different winding numbers, indicating the appearance of bipolar NHSE. The light yellow and dark yellow regions in Fig.~\ref{figs6} represent the twisting region. The light yellow region represents the appearance of twist winding in one band while the dark yellow region represents the simultaneous twisted loops in both bands.
      
   To characterize the localization properties of all eigenstates, we introduce the IPR of all eigenmodes (sIPR), defined as
   \begin{equation}
   	\text{sIPR} = \frac{1}{2N} \sum_{i=1}^{2N} \frac{\sum_{i=1}^{2N}|\Psi_n(\overrightarrow{r_i})|^4}{|\sum_{i=1}^{2N}|\Psi_n(\overrightarrow{r_i})|^2|^2}.
   	\label{eq29}
   \end{equation}
   The value of sIPR corresponds to the localization degree of eigenmodes in the system. sIPR close to 1 indicates a high degree of localization, while sIPR close to zero represents the eigenstates that are uniformly distributed through the system, similar to Bloch waves.
   
    As shown in Fig.~\ref{figs6}(b), we find that the system exhibits high localization around $|\phi_a|=|\phi_b|=\pi/2$. The reason is that when $|\phi_a|=|\phi_b|=\pi/2$, the imaginary part of the nonconservative coupling $\text{Im}(v)=v\sin(\phi_{a,b})$ is large, introducing significant non-Hermiticity into the system.

\section{Nonreciprocal energy transmission}
   \label{appendix3}
   \subsection{Nonreciprocal energy transmission in triangular plaquettes}
   In this section, we discuss the nonreciprocal energy transmission of the triangular lattice to further analyze the properties of the Hamiltonian shown in Eq.~(\ref{eq9}). The Hamiltonian of a single triangular lattice is given by
   \begin{equation}
   	H_{\text{cell}} = (\hat{a}_1^\dagger,\hat{b}_1^\dagger,\hat{a}_2^\dagger)\begin{pmatrix}
   		\delta & w & v\\
   		w^* & -\delta & u\\
   		v & u^* & \delta
   	\end{pmatrix}\begin{pmatrix}
   	    \hat{a}_1\\\hat{b}_1\\\hat{a}_2
   	\end{pmatrix}.
   	\label{eq12}
   \end{equation}
   To derive the scattering matrix of a single triangular lattice, we first write the corresponding Langevin equations in matrix form as
   \begin{equation}
   	\frac{d\vec{\nu}}{dt} = M\vec{\nu} + \sqrt{\Gamma_{\text{ex}}}\vec{\nu}_{\text{in}} + \sqrt{\Gamma_0}\vec{f}_{\text{in}},
   	\label{eq13}
   \end{equation}
   where the bases are $\vec{\nu} = (\hat{a}_1,\hat{b}_1,\hat{a}_2)^T$, $\vec{\nu}_{\text{in}} = (\hat{a}_1^{\text{in}},\hat{b}_1^{\text{in}},\hat{a}_2^{\text{in}})^T$, and $\vec{f}_{\text{in}} = (\hat{f}_1^{\text{in}},\hat{f}_1^{\text{in}},\hat{f}_2^{\text{in}})^T$. The damping rate matrix connected with the input field is $\sqrt{\Gamma_{\text{ex}}}=\text{Diag}[\sqrt{\gamma_{\text{ex}}},\sqrt{\gamma_{\text{ex}}},\sqrt{\gamma_{\text{ex}}}]$, and the intrinsic damping rate matrix is $\sqrt{\Gamma_{0}}=\text{Diag}[\sqrt{\gamma_{0}},\sqrt{\gamma_{0}},\sqrt{\gamma_{0}}]$, along with the noise field.
   The coefficient matrix is given by
   \begin{equation}
   \begin{aligned}
   	& M_{\text{cell}} = -i(H_{\text{cell}} + M_{\text{cell}}^{\gamma})\\
        &= -i\begin{pmatrix}
   		\delta & w & v\\
   		w^* & -\delta & u\\
   		v & u^* & \delta
   	\end{pmatrix}-i \begin{pmatrix}
   	-i\gamma_{\text{ex}}/2 & 0 & 0\\
   	0 & -i\gamma_{\text{ex}}/2 & 0\\
   	0 & 0 & -i\gamma_{\text{ex}}/2
   	\end{pmatrix}. 
    \end{aligned}
   	\label{eq14}
   \end{equation}
   To solve the Langevin equations Eq.~(\ref{eq13}) in the frequency domain, we introduce $\vec{\nu} = \frac{1}{2\pi} \int d\omega e^{-i\omega t} \vec{\nu(\omega)}$. Considering the intrinsic damping rate $\sqrt{\gamma_{0}} = 0$ (i.e., $\sqrt{\gamma_a} = \sqrt{\gamma_b} \approx \sqrt{\gamma_{\text{ex}}} \equiv \sqrt{\gamma}$) and substituting the solution into the input-output relation $\vec{\nu}_{\text{out}}(\omega) = \vec{\nu}_{\text{in}}(\omega) - \sqrt{\Gamma_{\text{ex}}}\vec{\nu}(\omega)$, where $\vec{\nu}_{\text{out}} = (a_1^{\text{out}},b_1^{\text{out}},a_2^{\text{out}})^T$ is the output operator, we obtain
   \begin{equation}
  	\vec{\nu}_{\text{out}}(\omega) = S(\omega)\vec{\nu}_{\text{in}}(\omega).
  	\label{eq15}
   \end{equation}
   The scattering matrix in frequency domain $S(\omega)$ can be written as
   \begin{equation}
   	S(\omega) = \mathbb{I} + \sqrt{\Gamma_{\text{ex}}}(i\omega\cdot \mathbb{I}+M_{\text{cell}})^{-1}\sqrt{\Gamma_{\text{ex}}},
   	\label{eq16}
   \end{equation}
   where $\mathbb{I}$ is the identity matrix. To simplify the calculation, consider $w \in \mathbb{R}$, $u \in \mathrm{i}\mathbb{R}$. On resonance, i.e. for $\omega = 0$, the matrix elements of $S(\omega)$ are given by 
   \begin{widetext}
   \begin{equation}
   	\begin{aligned}
   		S_{1,2} &= \frac{4i\gamma(-2\text{Im}(u)v+w(\gamma+2i\delta))}{4v^2(\gamma-2i\delta)+4\text{Im}(u)^2(\gamma+2i\delta)+(\gamma+2i\delta)(4w^2+\gamma^2+4\delta^2)},\\
   		S_{2,1} &= \frac{4i\gamma(2\text{Im}(u)v+w(\gamma+2i\delta))}{4v^2(\gamma-2i\delta)+4\text{Im}(u)^2(\gamma+2i\delta)+(\gamma+2i\delta)(4w^2+\gamma^2+4\delta^2)},\\
   		S_{1,3} &= \frac{4i\gamma(2\text{Im}(u)w+v(\gamma-2i\delta))}{4v^2(\gamma-2i\delta)+4\text{Im}(u)^2(\gamma+2i\delta)+(\gamma+2i\delta)(4w^2+\gamma^2+4\delta^2)},\\
   		S_{3,1} &= \frac{4i\gamma(-2\text{Im}(u)w+v(\gamma-2i\delta))}{4v^2(\gamma-2i\delta)+4\text{Im}(u)^2(\gamma+2i\delta)+(\gamma+2i\delta)(4w^2+\gamma^2+4\delta^2)},\\
   		S_{2,3} &= \frac{4\gamma(2wv-\text{Im}(u)(\gamma+2i\delta))}{4v^2(\gamma-2i\delta)+4\text{Im}(u)^2(\gamma+2i\delta)+(\gamma+2i\delta)(4w^2+\gamma^2+4\delta^2)},\\
   		S_{3,2} &= \frac{4\gamma(2wv+\text{Im}(u)(\gamma+2i\delta))}{4v^2(\gamma-2i\delta)+4\text{Im}(u)^2(\gamma+2i\delta)+(\gamma+2i\delta)(4w^2+\gamma^2+4\delta^2)}.
   	\end{aligned}  
   	\label{eq17}
   \end{equation}
    \end{widetext}
  In Eq.~\ref{eq17}, it is shown that the nonreciprocal energy flow exists only if $\text{Im}(u)v \neq 0$ which means that both the imaginary of $u$ and the nonconservative coupling strength $v$ are non-zero. Moreover, due to the connection between the NHSE and the nonreciprocity discussed in the paper, we find that the NHSE also exits only if $\text{Im}(u)v \neq 0$. Therefore, we consider the case where $u=i\delta/2$ in the paper.
   
\begin{figure}
   	\centering
   	\includegraphics[width=8.5cm]{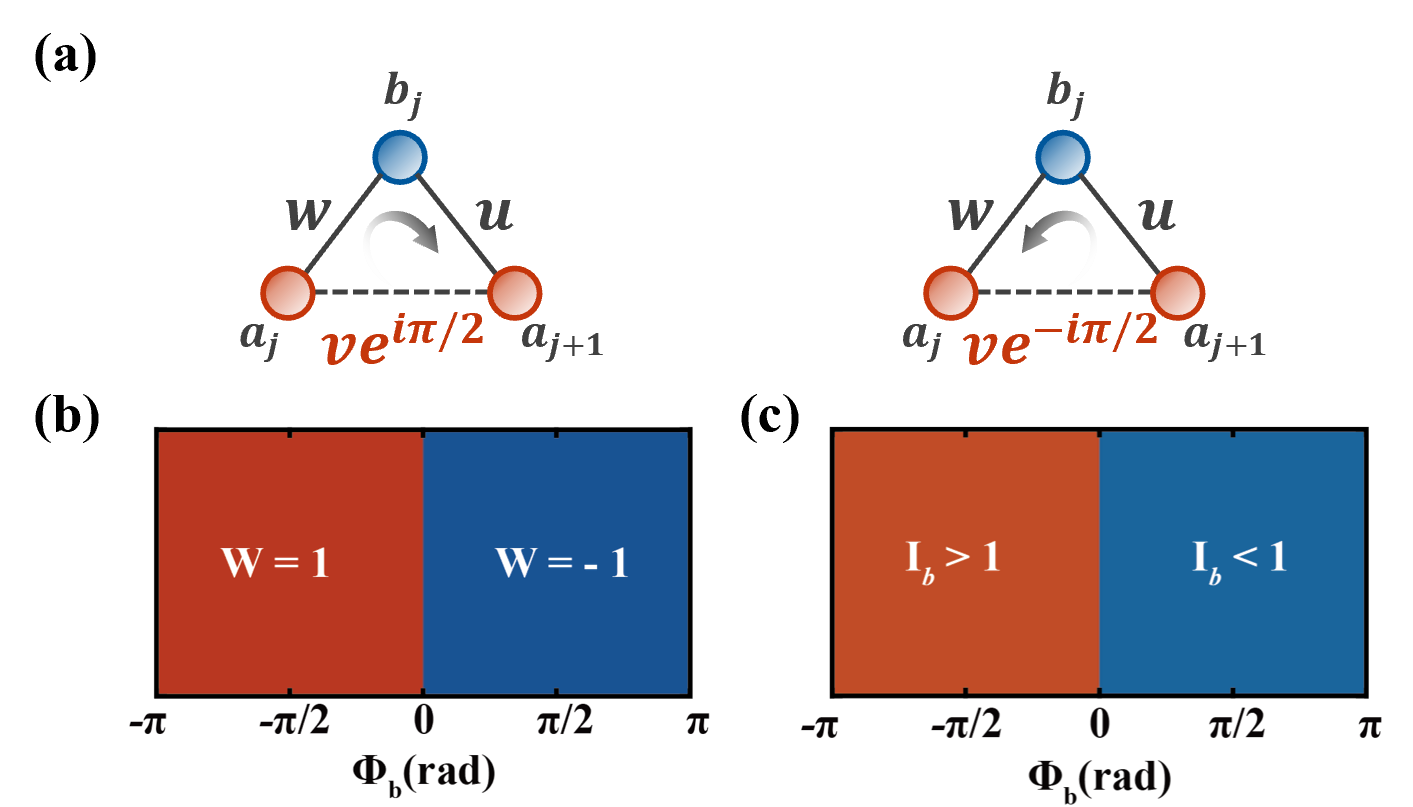}
   	\caption{(a) Schematic of the non-Hermitian triangular plaquettes with conservative couplings $w$ and $u$ (solid lines), and nonconservative coupling $v$ (dashed line). The gray arrows illustrate the direction of nonreciprocal flow in the triangular plaquette with different coupling phases. (b) Phase diagram of the spectral winding number for the phase transition of two energy bands in an upper triangular plaquette. The red region indicates $W=1$ and the blue region indicates $W=-1$. (c) Diagram of nonreciprocity ratios as a function of phases, with the input and output field connected to mode $b_1/b_N$. The red indicates the nonreciprocity ratio $I_b>1$, and the blue region indicates $I_b<1$. The size of the chain is $N=5$, and other parameters are $w/\delta = 1/2,\,u/\delta = i/2,\,v/\delta = 1/2$.}
   	\label{figs2}
   \end{figure}

   Specifically, Fig.~\ref{figs2}(a) shows that the energy transmission direction for $\phi_a = \pi/2$ is $a_1\rightarrow b_1 \rightarrow a_2 \rightarrow a_1$, forming an anticlockwise loop in a triangular lattice. For $\phi_a = -\pi/2$, the energy transmission direction is $a_1\rightarrow a_2 \rightarrow b_1 \rightarrow a_1$, forming a clockwise energy transmission loop in a single triangular lattice.

   Next, we extend the single triangular lattice to an upper triangular plaquette. The Hamiltonian is written as
   \begin{equation}
   	\begin{aligned}
    H_{\text{tri}} = &\sum_{i=1}^{N} [(\delta \hat{a}_i^\dagger \hat{a}_i - \delta \hat{b}_i^\dagger \hat{b}_i)+(w\hat{a}_i^\dagger \hat{b}_i+w^*\hat{b}_i^\dagger \hat{a}_i)]\\
   		&+\sum_{i=1}^{N-1} [v e^{i\phi_b}(\hat{b}_i^\dagger \hat{b}_{i+1}+\hat{b}_{i+1}^\dagger \hat{b}_i)+(u\hat{b}_i^\dagger \hat{a}_{i+1}+u^*\hat{a}_{i+1}^\dagger \hat{b}_i)].
   	\end{aligned}
   	\label{eq18}
   \end{equation}
    The scattering matrix of the chain can be written as
   \begin{equation}
   	S(\omega)_{\text{chain}} = \mathbb{I} + \sqrt{\Gamma_{\text{ex}}}(i\omega \cdot\mathbb{I}+M_{\text{tri}})^{-1}\sqrt{\Gamma_{\text{ex}}},
   	\label{eq19}
   \end{equation}
   where $M_{\text{tri}} = -iH_{\text{tri}}-iM_{\text{tri}}^{\gamma}$,
   \begin{widetext}
   \begin{equation}
   	M_{\text{tri}}^{\gamma} = \begin{bmatrix}
   		-i \gamma_{a_1}/2 & 0 & 0 & \cdots & 0 & 0 \\
   		0 & -i \gamma_{b_1}/2 & 0 & \cdots & 0 & 0 \\
   		0 & 0 & -i \gamma_{a_2}/2 & \cdots & 0 & 0\\
   		\vdots & \vdots & \ddots & \vdots & \vdots & \vdots\\
   		0 & 0 & \cdots & -i \gamma_{b_{n-1}}/2 & 0 & 0\\
   		0 &  0 & \cdots & 0 & -i \gamma_{a_n}/2 & 0 \\
   		0 & 0 & \cdots & 0 & 0 & -i \gamma_{b_n}/2 \\
   	\end{bmatrix},
   	\label{eq20}
   \end{equation}
   \begin{equation}
   	\sqrt{\Gamma_{\text{ex}}} = \begin{bmatrix}
   		\sqrt{\gamma_{a_1}} & 0 & 0 & \cdots & 0 & 0 \\
   		0 & \sqrt{\gamma_{b_1}} & 0 & \cdots & 0 & 0 \\
   		0 & 0 &  \sqrt{\gamma_{a_2}} & \cdots & 0 & 0\\
   		\vdots & \vdots & \ddots & \vdots & \vdots & \vdots\\
   		0 & 0 & \cdots & \sqrt{\gamma_{b_{n-1}}} & 0 & 0\\
   		0 &  0 & \cdots & 0 & 	\sqrt{\gamma_{a_n}} & 0 \\
   		0 & 0 & \cdots & 0 & 0 & 	\sqrt{\gamma_{b_n}} \\
   	\end{bmatrix}.
   	\label{eq21}
   \end{equation}
   \end{widetext}
   
  To analyze the connection between the nonreciprocity performance of the chain and the nonconservative coupling phases, we consider two different nonreciprocal transmission cases by selecting the leftmost mode $a_{1}$ ($b_{1}$) and rightmost mode $a_{N}$ ($b_{N}$) to connect the input and output fields, respectively. The nonreciprocity ratios for the two cases can be defined as $I_a  =T_{a\leftarrow}/T_{a\rightarrow}=|S_{1,2N-1}/S_{2N-1,1}|^2$ and $I_b =T_{b\leftarrow}/T_{b\rightarrow}= |S_{2,2N}/S_{2N,2}|^2$, which can be calculated by numerically solving the system scattering matrix $S(\omega)$. Using Eq.~(\ref{eq16}), we can obtain the nonreciprocity ratios for the two cases numerically.
   
   To simplify the calculation, consider the case where $w \in \mathbb{R},~u \in \mathrm{i}\mathbb{R}$. In the first scenario, we analyze nonreciprocal transmission by connecting the input field to the leftmost mode $a_1$ and the output field to the rightmost mode $a_N$. Assuming $\gamma_{a_1}=\gamma_{a_N}=\gamma_a$, the damping rate matrix connected with the input and output field is given by
   \begin{equation}
   	M_{\text{tri}}^{\gamma} = \begin{bmatrix}
   		-i \gamma_{a}/2 & 0 & 0 & \cdots & 0 & 0 \\
   		0 & 0 & 0 & \cdots & 0 & 0 \\
   		0 & 0 & 0 & \cdots & 0 & 0\\
   		\vdots & \vdots & \ddots & \vdots & \vdots & \vdots\\
   		0 & 0 & \cdots & 0 & 0 & 0\\
   		0 &  0 & \cdots & 0 & -i \gamma_{a}/2 & 0 \\
   		0 & 0 & \cdots & 0 & 0 & 0 \\
   	\end{bmatrix},
    \label{eq24a}
   \end{equation}
   \begin{equation}
   	\sqrt{\Gamma_{\text{ex}}} = \begin{bmatrix}
   		\sqrt{\gamma_{a}} & 0 & 0 & \cdots & 0 & 0 \\
   		0 & 0 & 0 & \cdots & 0 & 0 \\
   		0 & 0 &  0 & \cdots & 0 & 0\\
   		\vdots & \vdots & \ddots & \vdots & \vdots & \vdots\\
   		0 & 0 & \cdots & 0 & 0 & 0\\
   		0 &  0 & \cdots & 0 & 	\sqrt{\gamma_{a}} & 0 \\
   		0 & 0 & \cdots & 0 & 0 & 0\\
   	\end{bmatrix}.
   	\label{eq24b}
   \end{equation}
   Therefore, the nonreciprocity ratio for this scenario can be determined as 
   \begin{equation}
   	\begin{aligned}
   		I_a & = \left|\frac{S_{1,2N-1}}{S_{2N-1,1}}\right|^2 = \left|\frac{(vw+u\delta)(-uw+v\delta)^{N-2}}{(vw-u\delta)(uw+v\delta)^{N-2}}\right|^2.
   	\end{aligned}
   	\label{eq25}
   \end{equation}
   Similarly, in the second scenario, we analyze nonreciprocal transmission by connecting the input field to the leftmost mode $b_1$ and the output field to the rightmost mode $b_N$. Assuming $\gamma_{b_1}=\gamma_{b_N}=\gamma_b$, the damping rate matrix connected with the input and output fields in this case is given by
   \begin{equation}
   	M_{\text{tri}}^{\gamma} = \begin{bmatrix}
   		0 & 0 & 0 & \cdots & 0 & 0 \\
   		0 & -i \gamma_{b}/2 & 0 & \cdots & 0 & 0 \\
   		0 & 0 & 0 & \cdots & 0 & 0\\
   		\vdots & \vdots & \ddots & \vdots & \vdots & \vdots\\
   		0 & 0 & \cdots & 0 & 0 & 0\\
   		0 &  0 & \cdots & 0 & 0 & 0 \\
   		0 & 0 & \cdots & 0 & 0 & -i \gamma_{b}/2 \\
   	\end{bmatrix},
    \label{eq26a}
   \end{equation}
   \begin{equation}
   	\sqrt{\Gamma_{\text{ex}}} = \begin{bmatrix}
   		0 & 0 & 0 & \cdots & 0 & 0 \\
   		0 & \sqrt{\gamma_{b}} & 0 & \cdots & 0 & 0 \\
   		0 & 0 &  0 & \cdots & 0 & 0\\
   		\vdots & \vdots & \ddots & \vdots & \vdots & \vdots\\
   		0 & 0 & \cdots & 0 & 0 & 0\\
   		0 &  0 & \cdots & 0 & 	0 & 0 \\
   		0 & 0 & \cdots & 0 & 0 & \sqrt{\gamma_{b}}\\
   	\end{bmatrix}.
   	\label{eq26b}
   \end{equation}
   The nonreciprocity ratio of this case can be written as
   \begin{equation}
   	\begin{aligned}
   		I_b & = \left|\frac{S_{2,2N}}{S_{2N,2}}\right|^2 =\left|\frac{(uw-v\delta)^{N-1}}{(uw+v\delta)^{N-1}}\right|^{2}.
   	\end{aligned}
   	\label{eq27}
   \end{equation}
   We can find that in the presence of nonconservative couplings, the nonreciprocity ratios $I_a$ and $I_b$ become unequal to 1 only when $u \neq 0$, $w \neq 0$, and $\delta \neq 0$ .
   
   As shown in Fig.~\ref{figs2}(b) and (c), the red and blue regions in Fig.~\ref{figs2}(b) represent the closed-loop-like band structure with winding numbers $W=1$ and$-1$, respectively. In Fig.~\ref{figs2}(c), the red region indicates a nonreciprocity ratio $I_b>1$ while the blue region indicates $I_b<1$. We find that when $\text{sgn}(\phi_{b})>0$, the winding number of the corresponding band is $W=-1$ and the nonreciprocity ratio $I_{b}<1$. Conversely, when $\text{sgn}(\phi_{b})<0$, the winding number is $W=1$ and the nonreciprocity ratio $I_{b}>1$. Moreover, the disappearance of the interference induced by the upper and lower triangular plaquette leads to the disappearance of the twisting windings.
   
   \subsection{Time evolution with different initial states in the chain model}
\begin{figure}
		\centering
		\includegraphics[width=8cm]{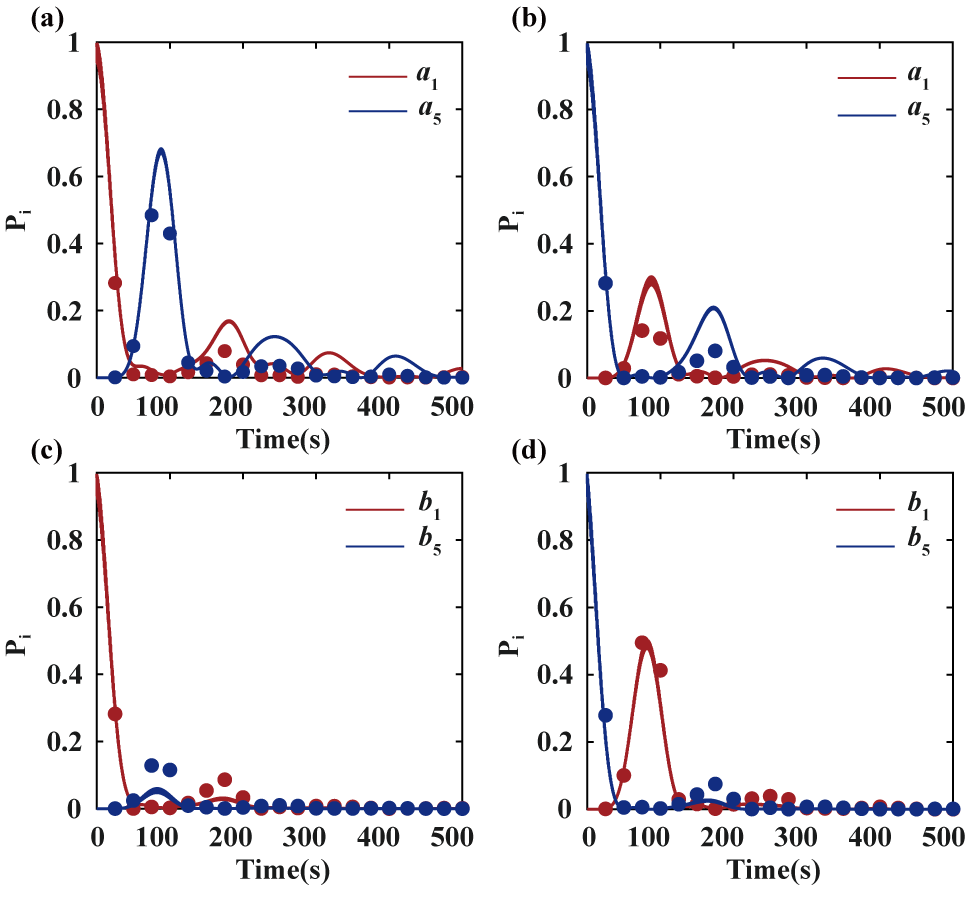}
		\caption{(a)(b) Time evolution of the energy in the first site $a_1$ (red) and the last site $a_{5}$ for energy initially stored in the leftmost site (a) and the rightmost site (b) of chain A. (c)(d) Time evolution of the energy in the first site $b_1$ (red) and the last site $b_{5}$ for energy initially stored in the leftmost mode (c) and the rightmost mode (d) of chain B. The curves are calculated from the original Hamiltonian, and the dots are obtained from the effective Hamiltonian. Other parameters are $\gamma/\delta = 5,\,\kappa = \gamma,\,\Delta_o = \Delta_q = 1.5\gamma,\,\Delta_a = 0.4\gamma,\,\Delta_b = -0.4\gamma,\,g = 0.1\sqrt{\gamma*|\Delta_o/\gamma+i/2|},\, w = \gamma/10,\,u = 3i/\gamma$, and the number of subcells is $N = 5$.}
		\label{figs5}
    \end{figure}

    Moreover, we discuss the time evolution of the effective Hamiltonian Eq.~(\ref{eq9}) with $N=5$ for various initial states. Using the Langevin equations Eq.~(\ref{eq13}) and ignoring the noise field, we can obtain the time evolution of the occupations of selected modes with different initial states. Considering $\phi_a = \phi_b = -\pi/2$, the eigenmodes predominantly localized on sites $a_j$ with minimal occupation on $b_j$ are localized at the right boundary, while the other group of eigenmodes is localized at the left boundary, see Fig. 3(a) and (b). In this case, energy can transmit from $a_1$ to $a_{5}$ with initial occupation at site $a_1$ [Fig.~\ref{figs5}(a)], while the transmission efficiency in the opposite direction is significantly smaller with initial occupation at $a_5$ [Fig.~\ref{figs5}(b)]. Moreover, energy can transmit from $b_{5}$ to $b_1$ with the initial occupation at $b_5$ [Fig.~\ref{figs5}(d)], while the opposite direction is significantly weaker [Fig.~\ref{figs5}(c)].
    \end{appendix}

\bibliography{ref-Bipolar-NHSE}

\end{document}